\documentclass[prl,aps,twocolumn,epsfig,showpacs]{revtex4}

\usepackage{amsmath}
\usepackage{amssymb}
\usepackage{graphicx}
\usepackage{dcolumn}
\usepackage{bm}
\usepackage[colorlinks=false,dvipdfm]{hyperref}
\usepackage{setspace}
\usepackage{amsfonts}
\usepackage{rotating}
\usepackage{makeidx}
\usepackage{amsthm}
\usepackage{graphics}

\begin{document}

\title{The Analytic Eigenvalue Structure of the 1+1 Dirac Oscillator \footnote{Chin. Phys. Lett. 37(9) 090303 (2020).}}
\author{Bo-Xing Cao}
%(曹博星)
\affiliation{Department of Physics, School of Science, Tianjin University, Tianjin
300072, China}
\author{Fu-Lin Zhang\footnote{Corresponding author. Email: flzhang@tju.edu.cn}}
%(张福林)
%\email[Corresponding author: ]{flzhang@tju.edu.cn}
\affiliation{Department of Physics, School of Science, Tianjin University, Tianjin
300072, China}
\date{\today}

\begin{abstract}
We study the analytic structure for the eigenvalues of the one-dimensional Dirac
oscillator, by analytically continuing its frequency on the complex plane. A
twofold Riemann surface is found, connecting the two states of a pair of
particle and antiparticle.
One can, at least in principle, accomplish the transition from a positive
energy state to its antiparticle state by moving the frequency continuously
on the complex plane, without changing the Hamiltonian after transition.
This result provides a visual explanation for the absence of a negative
energy state with the quantum number $n=0$.
\end{abstract}

%\begin{keyword}
% \sep  \sep
%\end{keyword}


%PACS 03.65.Pm { Relativistic wave equations
%PACS 03.65.Ge { Solutions of wave equations: bound states
%PACS 03.65.-w { Quantum mechanics

\maketitle
%
%\end{frontmatter}
%
%\linenumbers

%\section{Introduction}

Experimental studies of exceptional points in several systems \cite%
{PRL2001,PRL2011,Nature2016b,Nature2017,Nature2016} has resulted in the functions of a
complex variable no longer being limited to the role of abstract tool in the study of physics.
On the other hand, very recently,
theoretical research regarding the structure of the Riemann surface has been extended to a coupled quantum system,
and unexpected behaviors have been found. \cite{carl2016,alexander2018}
Specifically, in the system of two coupled harmonic oscillators, the presence of an eightfold Riemann surface structure of eigenvalues as a function of the complex coupling parameter has been discovered, which collapses to a fourfold structure in relation to the ground state.
By analytically continuing the system through the Riemann surface, and finally returning to the decoupling limit,
one can access the unconventional phases of both oscillators, which are originally found in the analytic continuation of the frequencies. \cite{bender1993}

In this work, we study the Dirac oscillator \cite{moshinsky1989,RevDO2010}  in one spatial dimension.
The motivation for this research is not only to extend the study of the harmonic oscillator to include its relativistic version,
but also based on the following considerations:
on the one hand, the one-dimensional Dirac oscillator can be derived by linearizing the quadratic form $E^2=m^2+p^2+m^2\omega^2 x^2- \beta m \omega$, as per Dirac's original approach to the proposal of his equation, \cite{DiracQM}  with $\omega $ being the frequency, and $\beta$ being one of the Dirac matrices.
Hence, it naturally has the property of a square root, which is a key cause of the structure of the Riemann surface in nonrelativistic harmonic
systems. \cite{bender1993,carl2016,alexander2018}
On the other hand, it can be regarded as a coupled quantum system, comprising a harmonic oscillator and a spin.
It may be interesting to re-examine the Riemann surface when one of the harmonic oscillators in the model studied in Refs. \cite{carl2016,alexander2018} is replaced by a spin.

The Dirac oscillator \cite{moshinsky1989,RevDO2010} is obtained from the free Dirac equation by the substitution $\bm{p}\rightarrow \bm{p}- \mathrm{i} \beta m \omega \bm{x}$, which has become the paradigm for the construction of covariant quantum models with a given well-determined nonrelativistic limit. \cite{PhysRevLett.111.170405}
It exhibits abundant algebraic properties \cite%
{lange1991,benitez1990,SymmDO,PhysRevC.69.024319,PhysRevA.80.054102}, and is
used in various branches of physics, such as nuclear physics \cite%
{PhysRevC.85.054617} and subnuclear physics \cite{PhysRevA.84.052102}.
Recently, there has been a growing interest in simulating the Dirac
oscillator in other physical systems, such as quantum optics \cite{PhysRevA.76.041801,PhysRevLett.99.123602,PhysRevLett.98.253005,AIPCP2010JC}
and classical microwave setups \cite{Sadurn__2010,PhysRevLett.111.170405}.
Progress both in terms of the experimental simulation of the Dirac oscillator, \cite{Sadurn__2010,PhysRevLett.111.170405} and the observation of exceptional points \cite{PRL2001,PRL2011,Nature2016b} via microwave setups,
have made it possible to experimentally observe the Riemann surface studied in this work.

By analytically continuing the frequency of the one-dimensional Dirac oscillator on the complex plane,
an unconventional spectrum arises, accompanying the conventional spectrum.
Such a characteristic is inherited from the nonrelativistic harmonic oscillator.
However, the difference here is that the conventional and unconventional states of the Dirac oscillator locate on different Riemann surfaces.
To be specific, although four eigenvalues, having the same quantum number $n$, can be expressed as a nested square-root function,
the absence of a symmetry of  $\omega \rightarrow -\omega$ in the Dirac oscillator leads to an
eigenfunction with a negative frequency, no longer belonging to the system.
Hence, the two conventional eigenvalues, i.e. of a particle and its antiparticle, should be considered to belong to two sheets of a square-root function of the frequency parameter $\omega$.
Moreover, the two unconventional eigenvalues belong to another square-root function.
On the twofold Riemann surface, one can move the Dirac oscillator from a positive energy state to its antiparticle state, by moving the system through a branch cut on the complex-frequency plane.
From this viewpoint, the disappearance of the branch point provides an interpretation for the absence of the negative energy state relating to $n=0$.
%

%\section{Harmonic Oscillator}\label{HO}

\textit{Harmonic Oscillator.}
Let us begin with a brief review of the analytical properties of the nonrelativistic harmonic oscillator.\cite{bender1993,carl2016,alexander2018}
Here, we show its eigenfunctions in both the conventional and unconventional spectra, based on which one may concisely solve the eigenvalue problem for the relativistic case.
The Hamiltonian of a nonrelativistic harmonic oscillator is given by
\begin{equation}  \label{HHO}
H=\frac {1}{2m} {p^2}+ \frac {1}{2} m\omega^2x^2 .
\end{equation}
In this work, we set the reduced Planck constant as $\hbar=1$.
In the asymptotic region $|x|\rightarrow +\infty $, $m\omega ^{2}x^{2}\gg |E|$,  the stationary Schr\"{o}dinger equation reads
\begin{equation}
\phi'' =m^{2}\omega ^{2}x^{2}\phi.
\end{equation}
Its approximate solutions can be written as
\begin{equation}
\phi =e^{f(x)},
\end{equation}
in which $f(x)=\pm \frac{1}{2}m\omega x^{2}$ is derived from
\begin{equation}\label{fprime}
 [ f^{\prime }(x)]^{2}=m^{2}\omega ^{2}x^{2}.
\end{equation}

Making the substitutions $\xi=\sqrt{m\omega}x$, $k=2E/\omega$ and $\phi (\xi )=h(\xi )e^{\mp \xi ^{2}/2}$, one obtains
\begin{equation}
\frac{d^{2}h}{d\xi ^{2}}\mp 2\xi \frac{dh}{%
d\xi }+(k\mp 1)h=0.
\end{equation}%
This will be the defining differential equation for the Hermite polynomials if $%
k=\pm (2n+1)$ with $n=0,1,2...$, which can then be represented as
\begin{equation}
h_{\pm n}(\xi )=(\pm 1)^{n}e^{\mp \xi ^{2}}\frac{\partial ^{n}}{\partial \xi
^{n}}e^{\pm \xi ^{2}}.
\end{equation}%
The plus sign corresponds to the conventional Hermite polynomials, $%
H_{n}(\xi )=h_{+n}(\xi )$, and the consequently conventional eigenfunctions
of the harmonic oscillator. We denote those with a minus sign as $h_{-n}(\xi )$%
, which leads to non-normalizable unconventional eigenfunctions. \cite{carl2016} The eigenvalues and corresponding eigenfunctions are then expressed as
\begin{eqnarray}\label{EigHO}
E_{n}^{\pm }=\pm (n+\frac{1}{2})\omega ,  \ \ \
\phi _{n}^{\pm }(\xi ) = h_{\pm n}(\xi )e^{\mp \frac{1}{2}\xi
^{2}}.
\end{eqnarray}

Clearly, the $\pm$ signs in the above results come from the square root of Eq. (\ref{fprime}).
Consequently, eigenvalues with the same quantum number, $n$, are merely different branches of a multivalued function of  complex $\omega$,
which can be expressed as
\begin{equation}
E_n (\omega)=\pm(n+\frac 12)\sqrt{\omega^2} =(n+\frac 12)\mathrm{e}^{\frac 12\mathrm{ln}(\omega^2)}.
\end{equation}
With the aid of the logarithmic function in the above expression, one may readily observe the connection structure of the function.

\begin{figure}
 \begin{center}
\includegraphics[width=8cm]{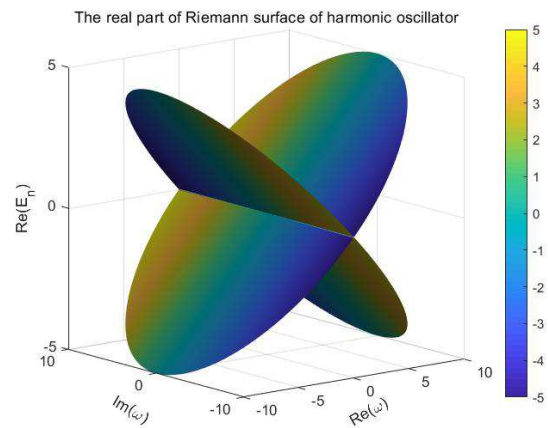}
\end{center}
\caption{
The real part of the Riemann surface for a one-dimensional harmonic oscillator with the quantum number $n=0$. The imaginary part is shown by color.
}\label{harmonicreal}
\end{figure}

In Fig. \ref{harmonicreal}, we show the real part of the Riemann surface for $n=0$.
The function is unchanged under $\omega \rightarrow - \omega$, which can be traced back to the symmetry of the Hamiltonian (\ref{HHO}).
However, the eigenvalues change sign, as the argument of $\omega$ runs continuously from $0$ to $\pi$.
There are two coalescing branch points at $\omega=0$, and the two associated branch cuts may be chosen along the imaginary $\omega$ axis.
Each additional $\pi$ on the argument of $\omega$ moves the system from one branch to another, changing the sign of the eigenvalues, although the Hamiltonian remains unchanged.
Simultaneously, the variable $\xi$ is changed into $i\xi$, leading to:
\begin{equation}
\phi^+_n( \xi)\rightarrow \phi^+_n(i\xi)= i^n\phi^-_n(\xi).
\end{equation}
Therefore, by virtue of such analytic continuation, we reach an unconventional state, from the starting point of a conventional one.
A similar process can likewise move the system from an unconventional state to a conventional one.

%\section{Dirac Oscillator} \label{DO}
%

\textit{Dirac Oscillator.} In this section, we study the one-dimensional version of the Dirac oscillator, whose Hamiltonian is given by
 \begin{equation}\label{HD}
\mathcal{H}= \alpha(p-i\beta m\omega x) +\beta m,
\end{equation}
where $\omega$ is a natural frequency, $\alpha$ and $\beta$ are the Dirac matrices,  and the light velocity is set to  $c=1$.
Here, we set the frequency $\omega$ to be positively real.
The Dirac matrices are conveniently defined in terms of the Pauli matrices:
\begin{equation}
\alpha=\sigma_y, \ \ \  \beta=\sigma_z.
\end{equation}

In order to observe its relation with the nonrelativistic harmonic oscillator, one can square the Hamiltonian (\ref{HD}), thereby obtaining
\begin{equation}\label{Hsq}
\mathcal{H}^2 =m^2+p^2+m^2\omega^2x^2- \sigma_z m\omega .
\end{equation}
and consequently,
\begin{eqnarray}\label{NR}
\frac{\mathcal{H}^2-m^2}{2m}= \frac{p^2}{2m}+ \frac{1}{2}m\omega^2x^2- \frac{1}{2} \sigma_z \omega.
\end{eqnarray}
In the nonrelativistic limit, $ m \gg \omega$ and $\mathcal{H}  \rightarrow m$, represent \textit{kinetic energy} $ \mathcal{H} -m \rightarrow {(\mathcal{H}^2-m^2)}/{2m}$.
The Dirac oscillator becomes a decoupled system, comprising a harmonic oscillator and a spin with a resonant frequency.
In addition, the leading term of the relativistic correction is proportional to $\omega^2 /m$.
Hence, the natural frequency $\omega$ also plays the role of a coupling constant.
On the other hand, the relations between Eqs. (\ref{Hsq}) and (\ref{NR}) enable one to solve the eigenvalue problem of $\mathcal{H}$, based on the results of the harmonic oscillator.
The process is described below.

Firstly, the eigenvalues and eigenfunctions of $\mathcal{H}^2$ can be obtained directly.
Denoting the eigenenergies of the Dirac oscillator as $\mathcal{E}$, the eigenvalues  of $\mathcal{H}^2$ are  $\mathcal{E}^2$.
Corresponding to each solution in (\ref{EigHO}), one can obtain
 \begin{eqnarray}
  \mathcal{E}^2=m^2 \pm 2 n  m  \omega.
\end{eqnarray}
and the two-component eigenstates
\begin{equation}
     %开始数学环境
     \renewcommand*{\arraystretch}{1.5}
\psi^+_n= \left(                 %左括号
  \begin{array}{c}   %该矩阵一共3列，每一列都居中放置
    a_n^+ \phi^+_n(\xi)\\  %第一行元素
    b_n^+\phi^+_{n-1}(\xi) \\  %第二行元素
  \end{array}
\right),                 %右括号
\psi^-_n = \left(                 %左括号
  \begin{array}{c}   %该矩阵一共3列，每一列都居中放置
    a_n^- \phi^-_{n-1}(\xi) \\  %第一行元素
   b_n^- \phi^-_{n}(\xi) \\  %第二行元素
  \end{array}
\right)   ,              %右括号
\end{equation}
where $\phi^\pm_{-1}(\xi)= \phi^\mp_{0}(\xi)$ and $\xi=\sqrt{m\omega}x$;
$a_n^\pm$ and  $b_n^\pm$ are free parameters, due to the double degeneration caused  by the resonance between the spin and harmonic oscillator in Eq. (\ref{NR}).
The states $\psi^+_0$ and $ \psi^-_0$ represent the same degenerate subspace.

Secondly, diagonalizing the Hamiltonian in the degenerate subspaces of $\mathcal{H}^2$, one can find the eigenfunctions of $\mathcal{H}$ as follows:
\begin{eqnarray}
\renewcommand*{\arraystretch}{1.5}
  {\Psi}_{\pm n}^+ =\left(                 %左括号
  \begin{array}{c}   %该矩阵一共3列，每一列都居中放置
     (m\pm\sqrt{m^2+2nm\omega})\phi^+_n(\xi) \\  %第一行元素
   2n\sqrt{m\omega}\phi^+_{n-1}(\xi)\\  %第二行元素
  \end{array}
\right),     \label{EigDO}\\
  \renewcommand*{\arraystretch}{1.5}
 \Psi_{\pm n}^- =\left(                 %左括号
  \begin{array}{c}   %该矩阵一共3列，每一列都居中放置
   2n\sqrt{m\omega}\phi^-_{n-1}(\xi) \\  %第一行元素
   (m\mp\sqrt{m^2-2nm\omega})\phi^-_n(\xi)\\  %第二行元素
  \end{array}
\right) ,    \label{EigDOM}
\end{eqnarray}
corresponding to the eigenvalues
 \begin{eqnarray}\label{EDO}
  \mathcal{E}^{\pm}_{\pm n} = \pm  \sqrt{m^2 \pm 2 n  m  \omega},
\end{eqnarray}
where the subscripts indicate the signs before the square roots, and the superscripts indicate those in the square roots.
When $n=0$,  $  {\Psi}_{- 0}^+ $ and $  \Psi_{+ 0}^-$ vanish,
and the eigenenergies $ \mathcal{E}^{\pm} = \pm m$  correspond to the states $  {\Psi}_{+ 0}^+$  and $  {\Psi}_{- 0}^-$, respectively.
The lower$/$upper component in $  {\Psi}_{+ 0}^+ /  \Psi_{- 0}^-$ is zero.
These results indicate that the conventional and unconventional states appear separately in $  {\Psi}_{\pm n}^+$ and $\Psi_{\pm n}^-$.
Hence, we refer to $\Psi_{\pm n}^-$ as the unconventional eigenfunctions of the Dirac oscillator, with  ${\Psi}_{\pm n}^+$  being conventional eigenfunctions.  \cite{radoslaw2001}

%\subsection{Analytic structure}\label{diracstructure}
%%\subsection{Spectra}

\begin{figure}
 \begin{center}
\includegraphics[width=8cm]{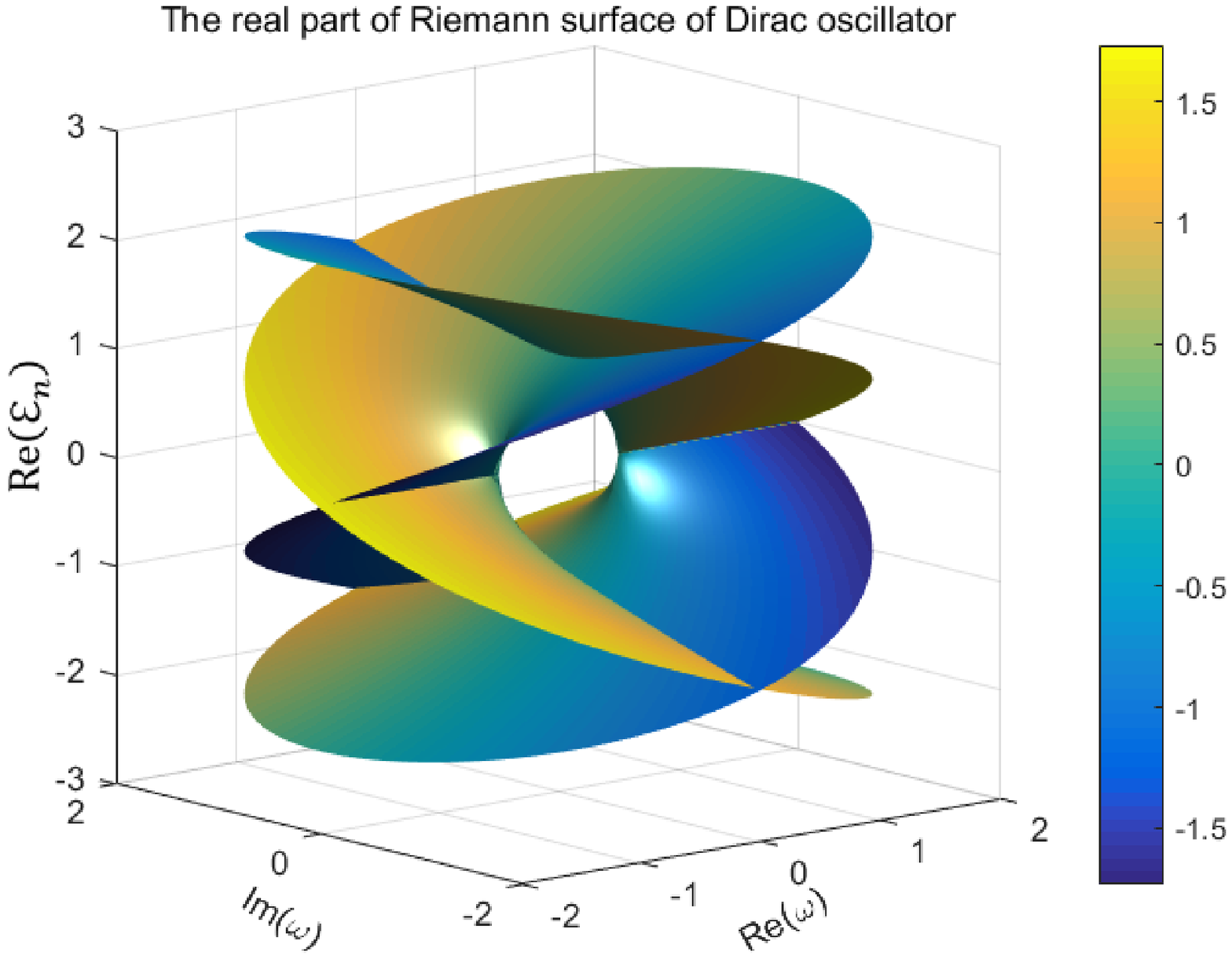}\\
 (a)\\
\includegraphics[width=8cm]{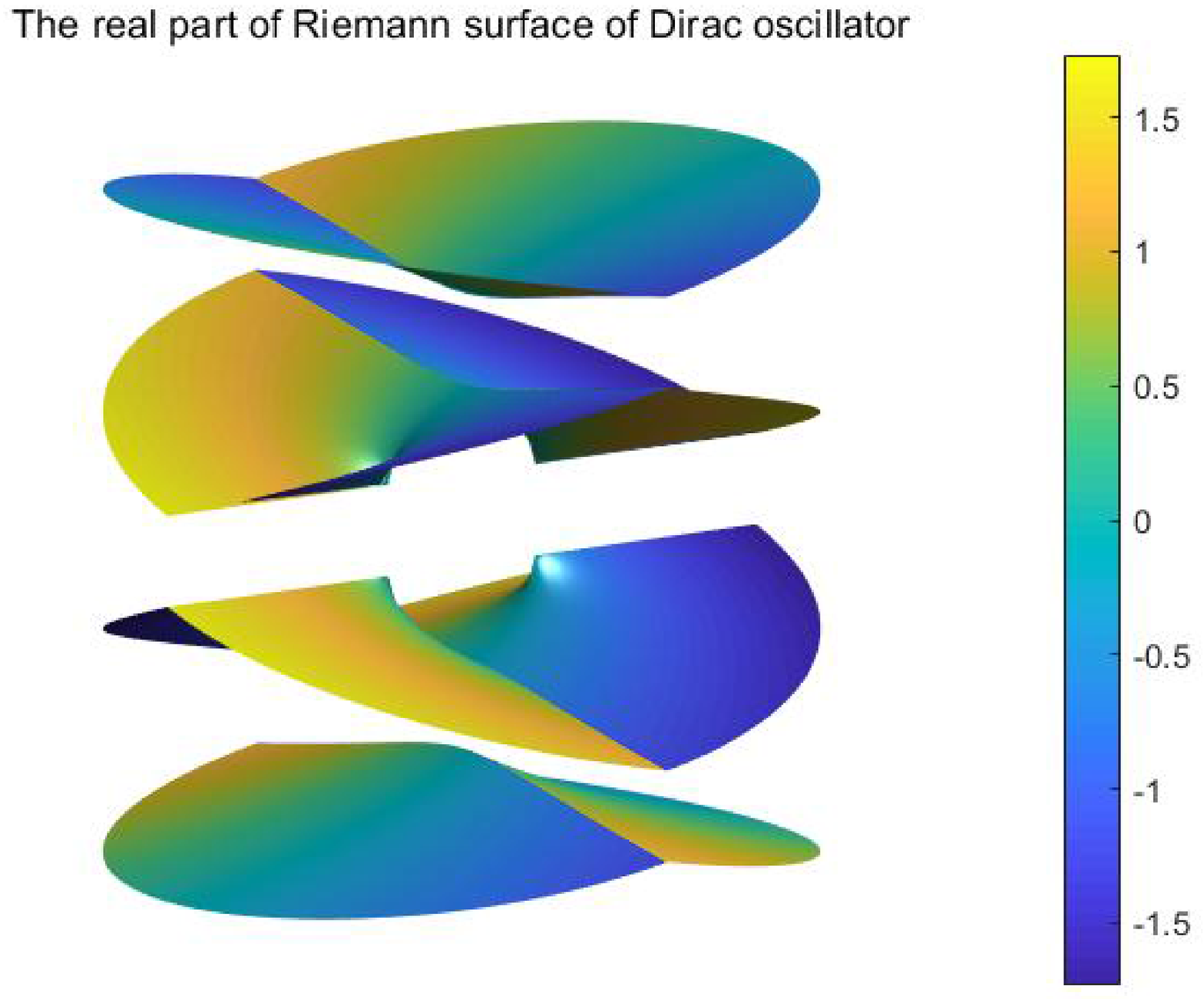}\\
(b)
\end{center}
\caption{
(a) The real part of Riemann surface of $\mathcal{E}_{  n}(\omega)$, where $m=1$, and $n=1$.
Shapes of the four sheets of the function are shown in (b).
The quantity of the imaginary part is shown in color.
}\label{FigEinDO}
\end{figure}

We can now analytically continue the frequency on the complex plane, and examine the analytic structure of the Dirac oscillator.
In this section, we show that,
although the eigenvalue (\ref{EDO}) with a fixed quantum number $n\neq0$  can be expressed as a nested square-root function,
the conventional states and unconventional states actually belong to two different square-root functions.

Since the the inner $\pm$ signs  (\ref{EDO}) originate from the square root to derive the eigenfunctions of the Harmonic oscillator,
it is natural to rewrite the eigenvalue as a nested square-root function
\begin{equation}
\mathcal{E}_{  n}(\omega)  \equiv \pm\sqrt{m^2 \pm 2 n m\sqrt{\omega^2}} .
\end{equation}
It has six square-root branch points, four occurring at $\omega=0$, and two at $\omega=\pm m/(2n)$.
The associated branch cuts at $\omega=0$ are chosen along the imaginary $\omega$ axis, and those at $\omega=\pm m/(2n)$ along the real $\omega$ axis.
These connect four sheets of the Riemann surface pairwise to one another.

A visualization of this surface and its four sheets is shown in Fig. \ref{FigEinDO}.
Let us take a brief look at the function.
We begin with a positive real $\omega=|\omega| >m/(2n)$ and $\mathcal{E}_{  n}(\omega)=\sqrt{m^2 + 2 n m\sqrt{\omega^2}}$, which locates on the top sheet.
Increasing the phase $\theta$ of $\omega=e^{i\theta} |\omega|$, one can enter the second sheet after passing through the positive-imaginary axis.
Next, running $\theta$ to the region of $(\pi,3\pi/2)$, one arrives the third sheet.
Subsequently, when $\omega$  reaches the fourth quadrant, $\mathcal{E}_{  n}(\omega)$ enters its fourth sheet, where $\mathcal{E}_{  n}(\omega)=-\sqrt{m^2 + 2 n m\sqrt{\omega^2}}$ when $\theta=2 \pi$.
If the phase $\theta$  continues to increase, the function will reenter the third, sencond, and first sheets in turn.
When $\theta=4 \pi$, $\mathcal{E}_{  n}(\omega)$ returns to the initial point.

An obvious feature of the above route is that only half the Riemann surface is reached by smooth  movement the $\omega$.
Starting from a posotive real $\omega$, one goes through the first and fourth sheets with $\mathrm{Re}(\omega)>0$, and the sencond and third ones with $\mathrm{Re}(\omega)<0$.
% $\mathrm{Im}(\omega)>0$
The four regions actually compose the Riemann surface of the two-valued function
\begin{equation}
\mathcal{E}^+_{  n}(\omega)\equiv \pm  \sqrt{m^2 + 2 n m\omega}.
\end{equation}
The start and end points of the above route represent two eigenvalues of the initial Hamiltonian, corresponding to two conventional eigenstates.
However, although the function on the negative-real axis equals two unconventional eigenvalues,
the state reached by verifying the frequency in a conventional eigenstate from  $ \omega $ to $- \omega $ is no longer an eigenstate of the initial Hamiltonian.
This difference from the Harmonic oscillator derives from the fact that the symmetry of $\omega \rightarrow -\omega$ is broken in the Dirac oscillator,
which is evident based on the linear terms in (\ref{Hsq}) and (\ref{NR}).
Therefore, the Riemann surface of the two-valued function $\mathcal{E}^+_{  n}(\omega)$ connects the two conventional states with the same quantum number $n$, and similarly one can find that the two-valued function
\begin{equation}
\mathcal{E}^-_{  n}(\omega) \equiv \pm \sqrt{m^2 - 2 n m\omega}.
\end{equation}
connects the two unconventional states.
Starting from one of the conventional states in expression (\ref{EigDO}), it is impossible to reach an unconventional state of the same Hamiltonian by smoothly varying the frequency $\omega$, and vice versa.

\begin{figure}
 \begin{center}
\includegraphics[width=8cm]{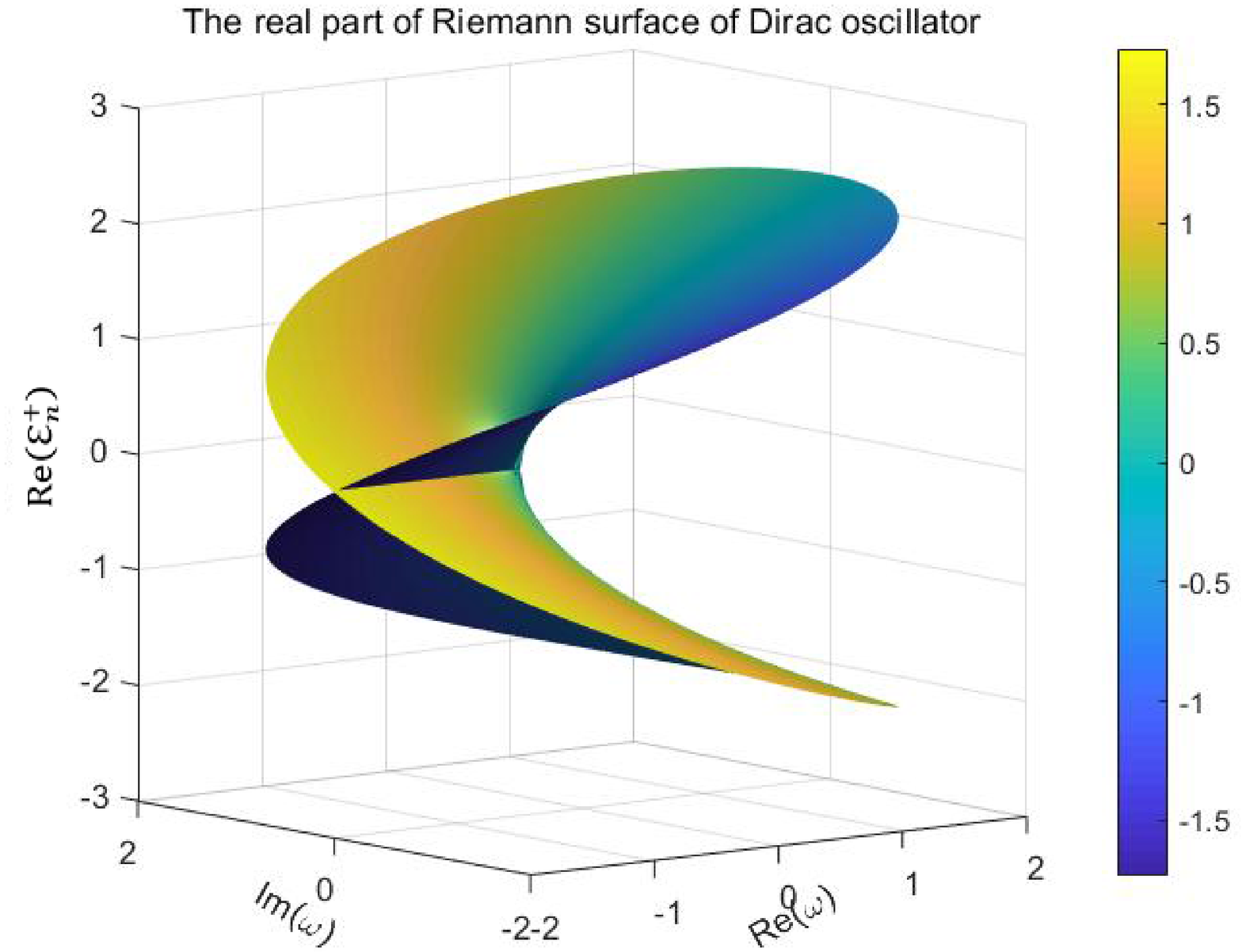}\\
 (a)\\
\includegraphics[width=8cm]{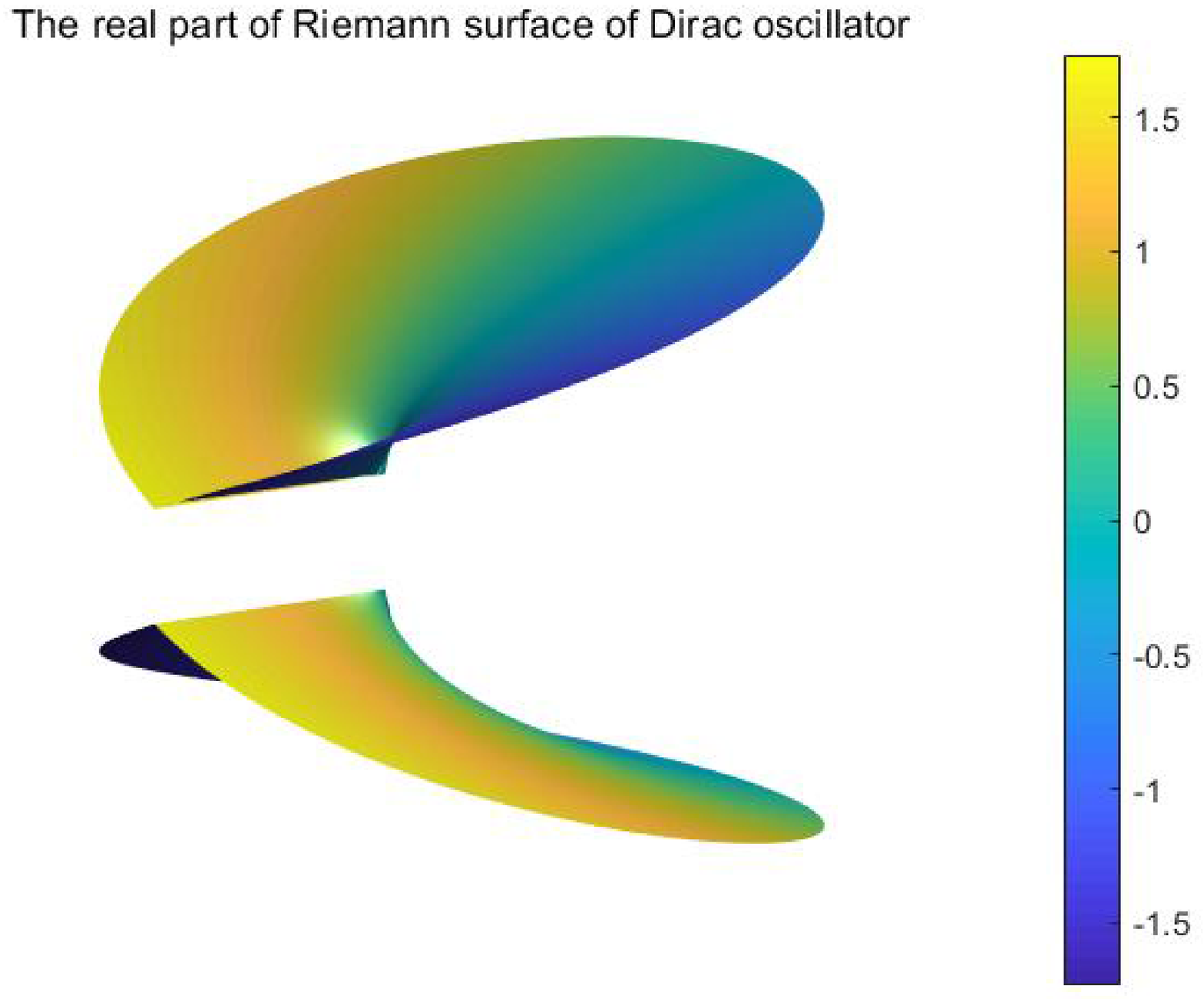}\\
(b)
\end{center}
\caption{
(a) The real part of Riemann surface of $\mathcal{E}^+_{  n}(\omega)$ with $m=1$ and $n=1$.
Shapes of the two sheets of  the function are shown in (b).
The amount of imaginary part is shown in color.
}\label{halfdiracreal}
\end{figure}

Based on the above considerations,
the multi-value functions connecting the eigenvalues are $\mathcal{E}^+_{  n}(\omega)$ and $\mathcal{E}^-_{  n}(\omega)$ rather than $\mathcal{E}_{  n}(\omega)$.
In order to highlight the connection structures, we take the conventional states as an example.
The results of the unconventional states can easily be obtained by simultaneously changing the two $\pm$ signs in (\ref{EDO}) and flipping the spin.
The Riemann surface of $\mathcal{E}^+_{  n}(\omega)$ is plotted in Fig. \ref{halfdiracreal}.
It has one square-root branch point at $\omega=- m/(2n)$ where $n\neq 0$, and the associated branch cut can be chosen along the negative-real axis.
Let the Dirac oscillator start from a positive energy state $\Psi _{+ n}^+ $ with $\mathcal{E}^+_{  n}(\omega)=+  \sqrt{m^2 + 2 n m\omega}$, we then run the argument $\theta$ of $\omega=\mathrm{e}^{\mathrm{i}\theta} |\omega|$ from $0$ to $2 \pi$.
When $|\omega|<  m/(2n)$, the system returns to its initial state  $\Psi _{+ n}^+ $.
When $|\omega|>  m/(2n)$, the system passes through the branch cut, and reaches its antiparticle state $\Psi _{- n}^+ $, with a negative energy $\mathcal{E}^+_{  n}(\omega)=-  \sqrt{m^2 + 2 n m\omega}$.
Here, $\Psi _{- n}^+ $ describes a state of an antiparticle with a positive energy $\sqrt{m^2 + 2 n m\omega}$.
According to the standard process in the textbooks \cite{RQM}, one can then find the operator for charge conjugation connecting the Dirac oscillator and its antiparticle system.
Specifically,   $\Psi _{+n}^{+,c}=\sigma_x (\Psi _{- n}^+)^*$  is an eigenstate of
 the antiparticle Hamiltonian $\mathcal{H}^c= \alpha(p+\mathrm{i}\beta m\omega x) +\beta m$, with an eigenvalue $\mathcal{E}^{+,c}_{  n}(\omega)=  \sqrt{m^2 + 2 n m\omega}$.

On the negative-real axis, the Hamiltonian in the degenerate subspaces of $\psi^+_n$ can be written as
\begin{eqnarray}
\mathcal{H}^+_n=  \left(                 %左括号
  \begin{array}{cc}   %该矩阵一共3列，每一列都居中放置
    m &  i \sqrt{m | \omega|} \\  %第一行元素
     i 2 n  \sqrt{m  |\omega|}  & - m \\  %第二行元素
  \end{array}
\right).
\end{eqnarray}
Under the similarity transformation, this is equivalent to
\begin{eqnarray}
\mathcal{H}^{+\prime}_n=  \left(                 %左括号
  \begin{array}{cc}    i \sqrt{2 n m | \omega|}  & m   \\
             m  & - i \sqrt{2 n m | \omega|}  \\  %第二行元素
  \end{array}
\right),
\end{eqnarray}
which is a special case of the $2\times 2$ PT-symmetric matrix Hamiltonian discussed in Ref. \cite{bender2003must}.
Here, we note that the time-reversal operator for the  $2\times 2$  systems in \cite{bender2003must}
is different from that for fermions, as described in more recent works.\cite{PhysRevA.98.022105,PhysRevA.99.062117}
In the transition between a state and its antiparticle state, the system goes through a region of broken PT symmetry on the negative-real axis,
as the eigenvalue becomes complex when $\theta=\pi$,
which may also be observed in the system comprising two coupled harmonic oscillators \cite{alexander2018}.
This PT symmetry breaking can be regarded as a relativistic effect, as it requires a large enough \textit{kinetic energy} that $ 2n |\omega| > m$.
The lower bound of $ |\omega|$ to break the PT symmetry increases with the quantum number $n$ decreases.
When $n = 0$, the bound becomes $+ \infty$.
This result provides a visual explanation for the absence of $\Psi _{- 0}^+ $.

%\section{Summary}\label{Conclu}

In summary, we have investigated the analytic structure of eigenvalues of the Dirac oscillator in one spatial dimension.
Here, we have four eigenfunctions, corresponding to an oscillator quantum number $n$, two of which are conventional states for a pair of particle and antiparticle, while the remaining two are unconventional.
The difference from the coupled-oscillator system \cite{carl2016,alexander2018} is that
conventional and unconventional states are separately connected by two distinct functions.
That is, the two conventional states with the same quantum number $n$ are connected by a twofold Riemann surface, for eigenvalues as a function of frequency on the complex plane;
this is also the case for the two unconventional states.
The system can be transitioned smoothly between the states by means of an analytic continuation in the frequency constant.
The transition occurs in a region with sufficient relativistic effects, where the PT symmetry is broken on the negative real axis.
Based on these results, the absence of a negative energy state with $n=0$ can be intuitively explained in terms of the disappearance of the branch point.

Further researches on this topic in several directions would be interesting.
Firstly, we look forward to an experimental verification of the analytic continuation studied in this work.
It would also be fascinating to consider the Berry phases \cite{Berry} acquired by the system when it moves on the Riemann surfaces.
Whether more elaborate structures arise from the Riemann surfaces in the two- or three-dimentional Dirac oscillators is also a natural question.
Finally, could we give a physical meaning to the unconventional states?
Or more specifically, is the negative energy of an unconventional state in nonrelativistic harmonic oscillators \cite{bender1993,carl2016,alexander2018}
actually a positive one, as per the negative energy in the Dirac equation?

%\section*{Acknowledgments}
%\begin{acknowledgments}
\textit{Acknowledgments.}  This work was supported by the National Natural Science Foundation of China (Grant Nos. 11675119, 11575125, and 11105097).
We thank Wu-Sheng Dai, Yun-Peng Liu and Wen-Ya Song for helpful discussions.
%\end{acknowledgments}

%\bibliography{diracoscillator}

\end{document}